\documentclass[12pt,a4paper]{article}

\usepackage[left=5em]{geometry}
\usepackage{amsmath}
\usepackage{graphicx}
\usepackage{amssymb}

\title{Stable Large-Scale Perturbations  In Interacting Dark-Energy Model}
\author{Cheng-Yi Sun\footnote{ddscy@163.com; cysun@mails.ucas.ac.cn}\
$^{,1}$, Rui-Hong Yue$^{2}$
\\
 {$^1$\small Institute of Modern Physics, Northwest University, Xian 710069, P.R.
 China}\\
{$^2$\small Faculty of Science, Ningbo University, Ningbo 315211,
P.R. China}}

\begin{document}
\maketitle
\begin{abstract}
It is found that the evolutions of density perturbations on the super-Hubble scales are unstable in the model with dark-sector interaction $Q$ proportional to the energy density of cold dark matter (CDM) $\rho_m$ and constant equation of state parameter of dark energy $w_d$. In this paper, to avoid the instabilities, we suggest a new covariant model for the energy-momentum transfer between DE and CDM. Then we show that the the large-scale instabilities of curvature perturbations can be avoided in our model in the universe filled only by DE and CDM. Furthermore, by including the additional components of radiation and baryons, we calculate the dominant non-adiabatic modes in the radiation era and find that the modes grow in the power law with exponent at the order of unit.
\end{abstract}



\section{Introduction}
We are convinced by the increasing observations
\cite{Supernova,WMAP,LSS} that the present universe is dominated by
the two components: dark energy (DE) \cite{dark energy1,dark energy2} and cold dark matter (CDM). Currently, the two
components are only indirected detected via the total gravitational effects. 
And then this produces an degeneracy: the two dark components might interact mutually without violating the observational constraints \cite{0702615,0801.3847}. And
furthermore it is found that an appropriate interaction can help to
alleviate the coincidence problem \cite{IDE0,IDE1}, namely why DE
and CDM are comparable in size exactly today \cite{dark energy2}.
Different interacting models of dark energy have been investigated
intensively \cite{IDE2,IDE3}.

Generally,  a model with interaction between DE
and CDM is described in the background by the two continuity equations
\begin{align}
  \label{CLBgM}
  \dot{\rho}_m+3H\rho_m&=Q,\\
  \label{CLBgD}
  \dot{\rho}_d+3H(1+w_d)\rho_d&=-Q,
\end{align}
where $Q$ denotes the phenomenological interaction term between DE
and CDM; $\rho_m$ and $\rho_d$ are the energy densities of CDM and
DE respectively; $w_d\equiv p_d/\rho_d$ is the equation of state
parameter of DE; $p_d$ is the pressure density of DE;
$H\equiv\dot{a}/a$ is the Hubble parameter; $a(t)$ is the scale
factor in the Friedmann-Robertson-Walker (FRW) metric; a dot denotes
the derivative with respect to the cosmic time $t$. In the note we
do not allow the phantom case $w_d<-1$. Owing to the lack of the
knowledge of micro-origin of the interaction, usually the
interaction term is parameterized in a simple form as \cite{IDE0}
\begin{equation}
  \label{Qgeneral}
  Q=3H(\alpha\rho_m+\beta\rho_d),
\end{equation}
where $\alpha$ and $\beta$ are positive constants. The interaction
term $Q$ would influence not only the background dynamics of the
universe, but also the growth of the perturbations of the
cosmological fluids \cite{0504571,Trodden,others}.

Recently, in Ref.\cite{0804.0232}, by modeling DE as a fluid with
constant $w_d>-1$, the authors investigated the evolution of the
linear density perturbations and found that the combination of
constant $w_d$ and the simple interaction form $Q$ given in
Eq.(\ref{Qgeneral}) leads to an instability: the curvature
perturbation on the super-Hubble scales blows up in the early
universe \cite{0804.0232}. The explicit models investigated in
Ref.\cite{0804.0232} included the two cases of $\beta=0$ and
$\alpha=\beta$ in Eq.(\ref{Qgeneral}). Further more, in
\cite{0901.3272}, it is concluded that the perturbations in the dark
energy become unstable for any model with constant $w_d>-1$ and
non-zero $\alpha$, no matter how small the parameter $\alpha$ is
made. In \cite{0807.3471,0901.3272,0901.1611}, the case of
$\alpha=0$ was surveyed and it was found that the instability can be
avoided if $\beta$ is made small enough. In \cite{0808.1646}, by
modeling DE as a quintessence field, the author found that the
instability can also be avoided even for the interaction $Q$
proportional to $\rho_m$.

Then it seems that a model with constant $w_d$ and non-zero
$\alpha$ in Eq.(\ref{Qgeneral}) would be ruled out as a viable interacting
model. To cure the curvature perturbations, the authors in Ref.\cite{1110.1807} suggested a covariant model for energy-momentum transfer between DE and CDM, and  showed that the instabilities of the curvature perturbation on the large scales can be avoided in the covariant model with constant $w_d$ and $Q=3\alpha H\rho_m$. In Ref.\cite{1110.1807}, the authors defined the effective EMTs of CDM and DE respectively as
\begin{align}
  \label{EMTensorEffCDM}
  T_{\text{em}}^{\mu\nu}&=\rho_mu_m^\mu u_m^\nu+p^{\text{eff}}_m(u_m^\mu u_m^\nu+g^{\mu\nu}),\\
  \label{EMTensorEffDE}
  T_{\text{ed}}^{\mu\nu}&=\rho_du_d^\mu u_d^\nu+p^{\text{eff}}_d(u_d^\mu u_d^\nu+g^{\mu\nu}),
\end{align}
where $p^{\text{eff}}_m\equiv-\alpha\rho_m$, $p^{\text{eff}}_d\equiv p_d+\alpha\rho_m$, and  $u_m^\mu$ and $u_d^\mu$ are the four velocities of CDM and DE respectively. The two effective EMTs are taken to be conserved \cite{1110.1807}
\begin{equation}
  \label{CLEff}
  {T_{\text{em}}^{\mu\nu}}_{;\nu}={T_{\text{ed}}^{\mu\nu}}_{;\nu}=0,
\end{equation}
and the Einstein equation are assumed to be \cite{1110.1807}
\begin{equation}
  \label{EinsteinEqEffEMT}
  G^{\mu\nu}=8\pi G(T_{\text{em}}^{\mu\nu}+T_{\text{ed}}^{\mu\nu}).
\end{equation}
Based on Eqs.(\ref{CLEff}) and (\ref{EinsteinEqEffEMT}), the evolution of the curvature perturbations were surveyed in \cite{1110.1807} and no instabilities  were found. The model defined in Eqs.(\ref{CLEff}) and (\ref{EinsteinEqEffEMT}) has a distinguishable feature that the total EMT of DE and CDM is not conserved at the perturbative level $(T^{\mu\nu}_{\text{m}}+T^{\mu\nu}_{\text{d}})_{;\nu}\neq0$, while, in other covariant models, the total EMT is always  taken to be conserved in order to match the Einstein tensor. Of course, in \cite{1110.1807}, the non-conserved total EMT does not cause problems since it is the total effective EMT that appears on the right-hand side of the Einstein equation and is guaranteed to be conserved by Eq.(\ref{CLEff}).

But it is still puzzling and discomforting that the total EMT of DE and CDM is not conserved. 
We try to solve the puzzlement by suggesting a new covariant model in this paper. We consider DE as a fluid with constant $w_d$ that is coupled to CDM via a covariant energy-momentum transfer, from which $Q=3\alpha H\rho_m$ can be reduced at the background level. We show that the total EMT of CDM and DE is conserved and the stabilities of the curvature perturbation on the large scales can be avoided in our new covariant model. 

This paper is organized as follows. In Sec.\ref{SecCovModel}, we 
display our new covariant model for the energy-momentum transfer in the dark sector.
In Sec.\ref{SecEvolving}, by assuming the universe filled only by DE and CDM, we
survey the evolution of the density perturbations and show that
instabilities on the large scales can be avoided. In Sec.\ref{SecDNAMode}, by considering the effects of
the radiation (photons and neutrinos) and baryons, we investigate the
dominant non-adiabatic mode in the radiation era and show that no
non-adiabatic modes blow up. Finally, conclusions and discussions
are given.

\section{New Covariant  Model for Interaction in Dark Sector } 
\label{SecCovModel}

Usually, the covariant form for energy-momentum transfer is taken to be
\cite{0804.0232,oldCovForm}
\begin{equation}
  \label{EMTAQ}
  \nabla_\nu T^{\mu\nu}_A=Q^\mu_A,
\end{equation}
where $A=m,d$ to denote CDM and DE respectively, and
the condition $\sum_A{Q^\mu_A}=0$ is imposed in order for the total energy-momentum tensor to
be conserved. By comparing Eqs.(\ref{CLEff}) and (\ref{EMTAQ}), we can get the covariant interacting terms of the model in Eq.(\ref{CLEff}) as \cite{1110.1807}
\begin{align}
  \label{myQm}
  Q^{\mu}_m&=\alpha[\rho_m(u^\mu_m u^\nu_m+g^{\mu\nu})]_{;\nu},\\
  \label{myQd}
  Q^\mu_d&=-\alpha[\rho_m(u^\mu_d u^\nu_d+g^{\mu\nu})]_{;\nu}.
\end{align}
Clearly, at the perturbation level, $Q^{\mu}_m\neq-Q^{\mu}_d$ and so the total EMT of DE and CDM  is not conserved.

Yet, motivated by Eqs.(\ref{myQm}) and (\ref{myQd}), we find that a new covariant model can be constructed as
\begin{align}
  \label{CLEMTm}
  {T^{\mu\nu}_\text{m}}_{;\nu}&=(\rho_mu_m^\mu u_m^\nu)_{;\nu}=-Q^{\mu}_d,\\
  \label{CLEMTd}
  {T^{\mu\nu}_\text{d}}_{;\nu}&=[\rho_du_d^\mu u_d^\nu+p_d(u_d^\mu u_d^\nu+g^{\mu\nu})]_{;\nu}=Q^{\mu}_d,
\end{align}
where $Q^\mu_d$ is given in Eq.(\ref{myQd}). Obviously, in this model the total EMT of DE and CDM is conserved. 
Indeed, this new model is just a usual type of Eq.(\ref{EMTAQ}) with the choice of the covariant energy-momentum transfer as 
\[
  Q^{\mu}_m=-Q^{\mu}_d=\alpha[\rho_m(u^\mu_d u^\nu_d+g^{\mu\nu})]_{;\nu}.
\]
Our model is similar to the case in \cite{9408025,9904120,9908026} where
\[
  Q^{\mu}_m=-Q^{\mu}_d=\beta(\phi)T^{\nu}_{m\nu}\nabla^\mu\phi,
\]
in the sense that the interaction is determined by the energy density of CDM  and the four velocity of DE.
It can be easily checked that Eqs.(\ref{CLBgM}) and (\ref{CLBgD})  with $Q=3\alpha H\rho_m$ can be deduced from Eqs.(\ref{CLEMTm}) and (\ref{CLEMTd}) at the background level, respectively. Notably, this model has  a similar feature to the one in Ref.\cite{1110.1807} that the global quantity  $H$ in $Q=3\alpha H\rho_m$ is explained to be a local quantity $u^{\nu}_{d;\nu}$.

Now together with the Einstein equation
\begin{equation}
  \label{EinsteinEq}
  G^{\mu\nu}=8\pi G(T^{\mu\nu}_{\text{m}}+T^{\mu\nu}_{\text{d}}),
\end{equation}
we can study the evolution of the curvature perturbation on the large scales to check whether the large-scale instabilities could be avoided.

\section{Evolution of Density Perturbations}
\label{SecEvolving}
In this section, we apply the new model suggested in the last section to survey the evolution of the density perturbations by modeling DE as a fluid with constant $w_d$. For simplicity, we consider a flat universe filled only by DE and CDM. We choose the conformal Newtonian gauge and then the perturbed FRW metric in the conformal is given by
\begin{equation}
  \label{FRWFirOrd}
  ds^2=a^2(\tau)[-(1+2\phi)d\tau^2+(1-2\psi)d\textbf{x}^2],
\end{equation}
where $\phi$ and $\psi$ denote the scalar perturbations. The Friedmann equation reads
\begin{equation}
  \label{FE}
  \mathcal{H}^2\equiv\left(\frac{a'}{a}\right)^2=\frac{8\pi G}{3}(\rho_m+\rho_d)
\end{equation}
Hereafter, primes denote the derivatives respect to $\tau$. With $Q=3\alpha H\rho_m$ and constant $w_d$, we can obtain the background evolutions of $\rho_m$ and $\rho_d$ from Eqs.(\ref{CLBgM}) and (\ref{CLBgD}) as 
\begin{align}
  \label{CDMBg}
  \rho_m&=\rho_{m0}a^{-3(1-\alpha)},\\
  \label{DEBg}
  \rho_d&=\rho_{d0}a^{-3(1+w_d)}+\Big(\frac{\alpha}{\alpha+w_d}\Big)\rho_{m0}a^{-3}(a^{-3w_d}-a^{3\alpha}).
\end{align}
In this paper, we use the subscript $0$ to denote the present value of the corresponding parameter and $a_0=1$.

\subsection{Evolving equations of Perturbations}

When the perturbed metric in Eq.(\ref{FRWFirOrd}) is considered, the
four velocities of CDM and DE are
\begin{equation}
  \label{4VelocityPert}
  u^\mu_m=a^{-1}\Big(1-\phi,\partial_iv_m\Big),\quad
  u^\mu_d=a^{-1}\Big(1-\phi,\partial_iv_d\Big),
\end{equation}
where $v_m$ and $v_d$ are the peculiar velocity potentials of CDM
and DE respectively. Usually, we define the volume expansion rates
of CDM and DE (in Fourier space) respectively as
\begin{equation}
  \label{theta}
  \theta_m=-k^2v_m,\quad \theta_d=-k^2v_d.
\end{equation}
We use $\delta\rho_m$, $\delta\rho_d$ and $\delta p_d$ to denote the
first-order perturbations of the corresponding parameters, and introduce two dimensionless first-order parameters as
\begin{equation}
    \label{delta}
    \delta_m=\frac{\delta\rho_m}{\rho_m},\quad
    \delta_d=\frac{\delta\rho_d}{\rho_d}.
\end{equation}
The total curvature perturbation on the constant-$\rho$
($\rho=\rho_m+\rho_d$) surface is defined as
\begin{equation}
  \label{curvPert}
  \zeta=-\psi-\mathcal{H}\frac{\delta\rho_m+\delta\rho_d}{\rho'_m+\rho'_d}.
\end{equation}
Here, following the analysis in Ref.\cite{0804.0232}, we take
\begin{equation}
  \label{deltaPd}
  \delta
  p_d=\delta\rho_d+(1-w_d)[3\mathcal{H}(1+w_d+\alpha\frac{\rho_m}{\rho_d})\rho_d]\frac{\theta_d}{k^2}.
\end{equation}
Then from Eq.(\ref{CDMBg}), we get two equations as
\begin{align}
  \label{dDeltaMdtau}
  \delta_m'-3(1-\alpha)\psi'+\theta_m-\alpha\theta_d&=0,\\
  \label{dThetaMdtau}
  \theta'_m+\mathcal{H}(1+3\alpha)\theta_m-k^2(1-\alpha)\phi&=\alpha\left[\theta'_d+\mathcal{H}(1+3\alpha)\theta_d-k^2\delta_m\right].
\end{align}
And from Eq.(\ref{DEBg}), we get other two equations as
\begin{align}
  \label{dDeltaDdtau2}
  \begin{split}
    \delta'_d+3\mathcal{H}(1-w_d-\alpha\frac{\rho_m}{\rho_d})\delta_d&+9\mathcal{H}^2(1-w_d)(1+w_d+\alpha\frac{\rho_m}{\rho_d})\frac{\theta_d}{k^2}\\
    &+3\alpha\mathcal{H}\frac{\rho_m}{\rho_d}\delta_m+(1+w_d+\alpha\frac{\rho_m}{\rho_d})(\theta_d-3\psi')=0,
  \end{split}\\
  \label{dThetaDdtau2}
  \begin{split}
    \theta_d'-2\mathcal{H}\theta_d-3\mathcal{H}\frac{\alpha(1-\alpha)\frac{\rho_m}{\rho_d}}{1+w_d+\alpha\frac{\rho_m}{\rho_d}}\theta_d&-k^2\phi\\
             &-k^2\frac{\alpha\frac{\rho_m}{\rho_d}}{1+w_d+\alpha\frac{\rho_m}{\rho_d}}\delta_m-k^2\frac{\delta_d}{1+w_d+\alpha\frac{\rho_m}{\rho_d}}=0.
  \end{split}
\end{align}

In the conformal Newtonian gauge, the first-order Einstein equations (\ref{EinsteinEq}) gives us \cite{9506072} 
\begin{align}
  \label{EinsteinEq00}
  &3\mathcal{H}\psi'+k^2\psi+3\mathcal{H}^2\phi=-4\pi Ga^2(\delta_m\rho_m+\delta_d\rho_d),\\
  \label{EinsteinEq0i}
  &k^2\psi'+k^2\mathcal{H}\phi=4\pi Ga^2[\rho_m\theta_m+(1+w_d)\rho_d\theta_d]\\
  \label{EinsteinEqii}
  &\psi''+\mathcal{H}(2\psi'+\phi')+(2\frac{a''}{a}-\mathcal{H}^2)\phi+\frac{k^2}{3}(\psi-\phi)=4\pi Ga^2\delta p_d\\
  \label{EinsteinEqij}
  &\psi-\phi=0.
\end{align}
Only two of the above equations are independent. Choosing any two of them and using Eqs.(\ref{dDeltaMdtau})-(\ref{dThetaDdtau2}), we can solve these evolving equations numerically if the initial conditions are given.

\subsection{Adiabatic Initial Conditions}

In the early universe, $a\ll1$, Eqs.(\ref{CDMBg}) and (\ref{DEBg})
indicate
\begin{equation}
  \label{rhoMOverRhoD}
  \frac{\rho_m}{\rho_d}\rightarrow-\frac{w_d+\alpha}{\alpha},
\end{equation}
and then, from Eq.(\ref{FE}), we have
\begin{equation}
  \label{initialMDHTau}
  \mathcal{H}=\frac{2}{1-3\alpha}\tau^{-1},\quad
  \tau=\frac{2}{(1-3\alpha)H_0}\sqrt{\frac{w_d+\alpha}{w_d\Omega_{m0}}}a^{\frac{1}{2}(1-3\alpha)},
\end{equation}
where
\[
  \Omega_{m0}\equiv\frac{8\pi G\rho_{m0}}{3H_0^2}.
\]
Here we adopt the adiabatic initial conditions to study the
evolution of the density perturbations on the super-Hubble scales
($k\ll aH$). To the lowest order in $k\tau$, we can set the
adiabatic conditions as
\begin{align}
  \label{initialPhiDis}
  \phi&=\psi=A_\phi=\text{Const.},\\
  \label{initialDelta}
  \delta_m&=\delta_d=-2A_\phi,\\
  \label{initialThetaM}
  \theta_m&=\theta_d=\frac{1-3\alpha}{3(1-\alpha)}k^2\tau A_\phi.
\end{align}

\subsection{Evolution of Curvature Perturbation on Large Scales}

\begin{figure}
\centering
\renewcommand{\figurename}{Fig.}
%\captionstyle{small}
\includegraphics[scale=1]{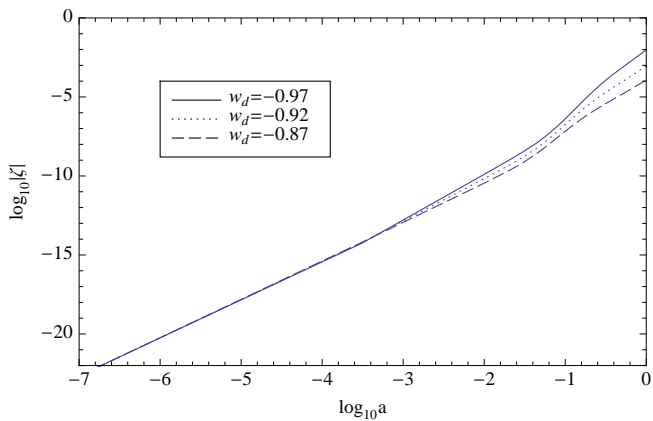}
\caption{$\log_{10}{|\zeta|}$ versus $\log_{10}{a}$ in the
interacting model for fixed $\Omega_{m0}=0.3$, $h=0.67$,
$k=1.5\times10^{-4}\text{Mpc}^{-1}$, $\alpha=10^{-3}$ and different
$w_d$.\label{FigZetaWd}}
\end{figure}

\begin{figure}
\centering
\renewcommand{\figurename}{Fig.}
%\captionstyle{small}
\includegraphics[scale=1]{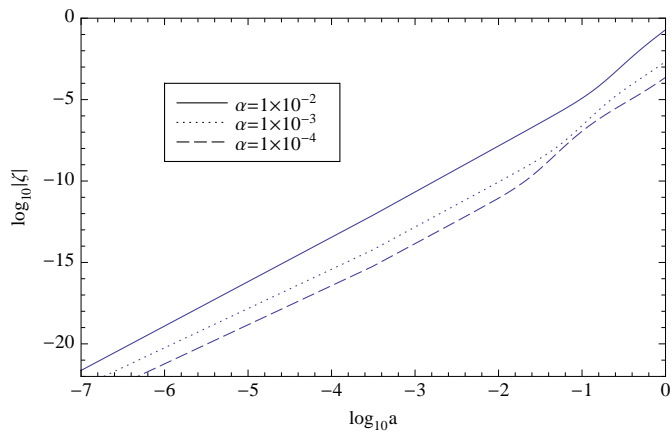}
\caption{$\log_{10}{|\zeta|}$ versus $\log_{10}{a}$ in the
interacting model for fixed $\Omega_{m0}=0.3$, $h=0.67$,
$k=1.5\times10^{-4}\text{Mpc}^{-1}$, $w_d=-0.94$ and different
$\alpha$.\label{FigZetaAlpha}}
\end{figure}

Now using the initial conditions in the last subsection, we can obtain the evolution of the scalar perturbations by solving Eqs.(\ref{dDeltaMdtau})-(\ref{EinsteinEq00}) and (\ref{EinsteinEqij}) numerically, and then get the evolution of the curvature perturbation $\zeta$ defined in Eq.(\ref{curvPert}). We display the results in Fig.\ref{FigZetaWd} and Fig.\ref{FigZetaAlpha}. Here we have taken $A_\phi=10^{-25}$ and $a_0=1$, and fixed the values of some parameters: $\Omega_{m0}=0.3$, $k=1.5\times10^{-4}\text{Mpc}^{-1}$, $H_0=100h\;\text{km}\;\text{sec}^{-1}\text{Mpc}^{-1}$ and $h=0.67$. The evolving curves in Fig.\ref{FigZetaWd} and Fig.\ref{FigZetaAlpha} manifest the regular growth in power law and no instabilities occur.

\section{Dominant Non-adiabatic Modes}
\label{SecDNAMode}

In the last section, by solving the evolving equations numerically with the adiabatic initial conditions, we show that instabilities of the curvature perturbation on the large scales are removed in our new covariant model. But the conclusion is obtained in the universe filled only by CDM  and DE. In this section, we consider the effects of baryons, photons and neutrinos by calculating the dominant non-adiabatic modes deep in the radiation era. If the dominant non-adiabatic modes do not grow rapidly, we believe that the instabilities can be avoided even when radiations and baryons are involved in the universe.

The $A$-fluid EMT including perturbations is
taken to be
\begin{equation}
  \label{EMTA}
  T^{\mu\nu}_A=(\rho_A+p_A)u^{\mu}_Au^{\nu}_A+p_Ag^{\mu\nu}+\pi^{\mu\nu}_A,
\end{equation}
where $u^\mu_A$ is the four velocity
\[
  u^{\mu}_A=a^{-1}\Big(1-\phi,\partial_iv_A\Big).
\]
We have allowed an anisotropic shear perturbation $\pi^{\mu\nu}_A$,
and $A=m,d,b,\gamma,\nu$ to denote the corresponding parameters of
CDM, DE, baryons, photons and neutrinos. Following Ref.\cite{0804.0232}, we take
$\pi_m^{\mu\nu}=\pi_d^{\mu\nu}=\pi_b^{\mu\nu}=\pi_\gamma^{\mu\nu}=0$ and 
\begin{equation}
  \label{piNeutrino}
  \pi_\nu^{0\mu}=0,\quad
  \pi^{ij}_\nu=a^{-2}\Big(\partial_i\partial_j-\frac{1}{3}\delta^{ij}\big)\pi_\nu.
\end{equation}
For DE and CDM, the continuity equations (\ref{CLEMTm}) and (\ref{CLEMTd}) still hold.

Early in the radiation era, the Friedmann equation reads
\begin{equation}
  \label{FERD}
  \mathcal{H}^2a^{-2}=\frac{8\pi G}{3}(\rho_\gamma+\rho_\nu)=\frac{8\pi
  G}{3}\rho_{r0}a^{-4}.
\end{equation}
Here we use the subscript $r$ to denote the corresponding parameter of radiation which consists of photons and neutrinos.
Then we have
\begin{equation}
  \label{initialRDHTau}
  a=\sqrt{\Omega_{r0}}H_0\tau,\quad \mathcal{H}=\tau^{-1}, \quad
  \Omega_{r0}\equiv\frac{8\pi G\rho_{r0}}{3H^2_0}.
\end{equation}
In the radiation era, the perturbed Einstein equations give us that
\begin{align}
  \label{EinsteinEqRD00}
  &3\tau^{-1}\psi'+k^2\psi+3\tau^{-2}\phi=-4\pi Ga^2\Big(\delta\rho_m+\delta\rho_d+\sum_A\delta\rho_A\Big),\\
  \label{EinsteinEqRD0i}
  &k^2(\psi'+\tau^{-1}\phi)=4\pi Ga^2\Big[\rho_m\theta_m+(1+w_d)\rho_d\theta_d+\sum_A(\rho_A+p_A)\theta_A\Big],\\
  \label{EinsteinEqRDii}
  &\psi''+2\tau^{-1}\psi'+\tau^{-1}\phi'-\tau^{-2}\phi+\frac{k^2}{3}(\psi-\phi)=4\pi
    Ga^2\Big(\delta p_d+\sum_A\delta p_A\Big),\\
  \label{EinsteinEqRDij}
  &\psi-\phi=8\pi G\pi_\nu,
\end{align}
where $A$ runs over $b,\gamma,$ and $\nu$.

For CDM and DE, the perturbed continuity equations are given by
Eqs.(\ref{dDeltaMdtau})-(\ref{dThetaDdtau2}). For baryons, the
perturbed continuity equations (in Fourier space) are
\cite{0804.0232}
\begin{align}
  \label{dDeltaBdtau}
  \delta_b'&=-\theta_b+3\psi',\\
  \label{dThetaBdtau}
  \theta_b'&=-\mathcal{H}\theta_b+k^2\phi,
\end{align}
and for photons \cite{0804.0232}
\begin{align}
  \label{dDeltaGammadtau}
  \delta_\gamma'&=-\frac{4}{3}\theta_\gamma+4\psi',\\
  \label{dThetaGammadtau}
  \theta_\gamma'&=\frac{1}{4}k^2\delta_\gamma+k^2\phi,
\end{align}
and for neutrinos \cite{0804.0232}
\begin{align}
  \label{dDeltaNudtau}
  \delta_\nu'&=-\frac{4}{3}\theta_\nu+4\psi',\\
  \label{dThetaNudtau}
  \theta_\nu'&=\frac{1}{4}k^2\delta_\nu+k^2\phi-k^2\sigma_\nu,\\
  \label{dSigmaNudtau}
  \sigma_\nu'&=\frac{4}{15}\theta_\nu,
\end{align}
where $\theta_A\equiv-k^2v_A$ for $A=b,\gamma,\nu$, and
$\sigma_\nu\equiv2k^2\pi_\nu/[3a^2(\rho_\nu+p_\nu)]$

In order to find the dominant non-adiabatic modes, we assume a
leading-order power law for perturbations \cite{0804.0232}
\begin{equation}
  \label{defineNonAdiaMod}
  \psi=A_\psi(k\tau)^{n_\psi},\ \phi=A_\phi(k\tau)^{n_\phi},\
  \delta_A=B_A(k\tau)^{n_A},\ \theta_A=C_A(k\tau)^{s_A},\
  \sigma_\nu=D_\nu(k\tau)^{n_\sigma}.
\end{equation}
Here the subscript $A=m,d,b,\gamma$, and $\nu$ denotes the
corresponding parameter of CDM, DE, baryons, photons and neutrinos,
respectively. To the leading order in $k\tau$, the equations
(\ref{dDeltaMdtau})-(\ref{dThetaDdtau2}) and
(\ref{EinsteinEqRD00})-(\ref{dSigmaNudtau}) may be solved, in terms
of $\psi$:
\begin{align}
  \label{DisplayPhiPsi}
  \phi&=J\psi,\quad J=1-\frac{16R_\nu}{5(n_\psi+2)(n_\psi+1)+8R_\nu},\\
  \label{DisplayDeltaGammaPsi}
  \delta_\gamma&=\delta_\nu=4\psi,\quad \delta_b=3\psi,\\
  \label{DisplayThetaGammaPsi}
  \theta_\gamma&=\theta_\nu=\theta_b=\frac{J+1}{n_\psi+1}k^2\tau\psi,\quad \sigma_\nu=\frac{4}{15}\frac{\tau\theta_\nu}{n_\psi+2}\\
  \label{DisplayDeltaPsi}
  \delta_m&=3(1-\alpha)\psi,\\
  \label{DisplayDeltaDPsi}
  \delta_d&=\frac{2\Omega_{r0}^{(1-3\alpha)/2}}{\alpha\Omega_{m0}H_0^{1+3\alpha}}(w_d+\alpha)(n_\psi+J+2)\frac{\psi}{\tau^{1+3\alpha}},\\
  \label{DisplayThetaDCDMPsi}
  \theta_d&=-\frac{n_\psi+2}{9(1-w_d)(1-\alpha)}k^2\tau\delta_d,\quad   \theta_m=\alpha\theta_d,
\end{align}
where $R_\nu\equiv\rho_\nu/(\rho_\gamma+\rho_\nu)$ and
\begin{equation}
  \label{DisplayNpsi}
  n_\psi=\frac{-3w_d\pm\sqrt{9w_d^2+12w_d-20}}{2},
\end{equation}
From Eq.(\ref{DisplayDeltaDPsi}), we should require
\[
  \text{Re}[n_\psi]\ge1+3\alpha.
\]
in order for the modes to be well behaved as $k\tau\rightarrow0$.
For $w_d\sim-1$, this leads to
\[
  -\frac{3}{2}w_d\ge1+3\alpha\Rightarrow\alpha\lesssim\frac{1}{6}.
\]

Correspondingly, the total curvature perturbation $\zeta$ is defined
as
\[
  \zeta=-\psi-\mathcal{H}\frac{\sum_A\delta\rho_A}{\sum_A\rho_A'}
\]
where $A$ runs over $m,d,\gamma,\nu$ and $b$. Then $\zeta$ can be
expressed in terms of $\psi$ as
\begin{equation}
  \label{zetaRDLeadingOrder}
  \zeta=-\frac{1}{2}(n_\psi+J+2)\psi.
\end{equation}
For $w_d\sim-1$, $n_\psi$ is a complex number and
\[
  \text{Re}[n_\psi]\sim\frac{3}{2}.
\]
So the dominant non-adiabatic modes grows in a power law with
exponent at the order of unit and no instabilities are present.

\section{Conclusions and Discussions}

It was found in \cite{0804.0232} that the evolution of the density perturbations is unstable in the model with constant $w_d$ and $Q$ proportional to $\rho_m$. To cure it, a covariant model for energy-momentum transfer in dark sector was suggested in \cite{1110.1807}. Yet, the model in \cite{1110.1807} seems to be ``anomalous", since it has a puzzling and discomforting feature that the total EMT of DE and CDM is not conserved. In order to remove the ``anomaly", in this paper, we have suggested a new covariant model in which the total EMT of DE and CDM is conserved. By choosing $Q=3\alpha H\rho_m$ and using the new model, we survey the evolution of density perturbations in a universe filled by DE  and CDM, and show that the large-scale instabilities can be avoided. Furthermore, we calculate the dominant non-adiabatic modes in the radiation era by including the other components of radiation and baryons in the universe, and find that the dominant non-adiabatic modes grow in the power law with exponent at the order of unit. This makes us believe that the instabilities can also be avoided even in a universe filled by radiation, matter and DE.

Qualitatively, it is easy to understand why the instabilities can be avoided in our new model. In the model in \cite{0804.0232}, it was shown that the coupling term $Q$ leading the instability-driving term in the perturbed continuity equation for $\theta_d$ during the radiation era
\begin{equation}
  \label{oldCovaModelInst}
  \theta_d'\sim\frac{\alpha}{1+w_d}\frac{\rho_m}{\rho_d}\mathcal{H}\theta_d\simeq-\frac{w_d+\alpha}{1+w_d}\mathcal{H}\theta_d.
\end{equation}
If $w_d$ is close to $-1$, this term becomes very large and causes the rapid growth of $\theta_d$ during the radiation era. Nevertheless, in the model in \cite{1110.1807}, it was shown that the corresponding driving term in the continuity equation for $\theta_d$ during the radiation era becomes
\begin{equation}
  \label{newCovaModelInst}
  \theta_d'\sim\frac{\alpha}{1+w_d+\alpha(\rho_m/\rho_d)}\frac{\rho_m}{\rho_d}\mathcal{H}\theta_d\simeq-\frac{w_d+\alpha}{1-\alpha}\mathcal{H}\theta_d.
\end{equation}
Clearly, this term does not cause instabilities even for $w_d\sim-1$. So it is believed that the instabilities can be avoided in the model in \cite{1110.1807}. In fact, the continuity equation for $\theta_d$ in our new model, Eq.(\ref{dThetaDdtau2}), are same to the one in \cite{1110.1807}. Then the analysis based on Eq.(\ref{newCovaModelInst}) also apply here. So we believe that the instabilities can be avoided in our new model, as in the model in \cite{1110.1807}.

\section*{Acknowledgments}
This work has been supported in part by the NNSF of China under
Grant No.11147017, the Research Fund for the Doctoral Program of
Higher Education of China under Grant No.20106101120023, the NNSF of
China under Grant No.11275099, and the Natural Science Foundation of the
Shaanxi Province under Grand No.2011JQ1002.

\end{document}